\documentclass[%
 reprint,
 amsmath,amssymb,
 aps,
]{revtex4-2}
 
\usepackage{graphicx}
\usepackage{dcolumn}
\usepackage{bm}
\usepackage{natbib}
\setcitestyle{super,sort,compress,comma}

\usepackage[colorlinks]{hyperref}
\usepackage[nameinlink]{cleveref} 
\usepackage{titlesec}
\usepackage{booktabs,tabularx}
\usepackage{enumitem}
\usepackage{url}
\usepackage{microtype}
\usepackage[english=usenglishmax]{hyphsubst}
\usepackage{subfigure}
\usepackage[left]{lineno}
\usepackage{xcolor}
\usepackage{listings}
\usepackage{verbatim}
\usepackage[T1]{fontenc}
\usepackage{enumitem}
\usepackage{makecell}
\usepackage{xr}
\usepackage{float}
\usepackage{comment}
\usepackage{hyperref}

\makeatletter
\renewcommand{\fnum@figure}{\textbf{\figurename~\thefigure}}
\def\@caption@fignum@sep{ }
\makeatother

\makeatletter
\renewcommand{\fnum@table}{\textbf{\tablename~\thetable}}
\makeatother

\renewcommand{\figurename}{\textbf{Fig.}}
\renewcommand{\tablename}{Tab.}

\setlist[enumerate]{itemsep=0.5em, topsep=0.5em, partopsep=0em, parsep=0em}

\hypersetup{
    colorlinks=true,
    linkcolor=blue,     
    citecolor=blue, 
    urlcolor=blue   
}

\smallskip

\titleformat*{\subsubsection}{\scriptsize\bfseries}

\titlespacing*{\section}{0pt}{0.1\baselineskip}{0.2\baselineskip}

\titlespacing*{\section}
{0pt}{2.5ex}{2.5ex}
\titlespacing*{\subsection}
{0pt}{2ex}{2ex}
\titlespacing*{\subsubsection}
{0pt}{2ex}{2ex}

\begin{document}

\title{Machine-Learning Potentials for sodium-potassium chloride mixtures: Predicting thermophysical properties and phase behavior of multicomponent salts}

\author{Karim Zongo$^{1, *}$}
\author{Hao Sun$^{1}$}
\author{Zijian Meng$^{1}$}
\author{Christopher Maxwell$^{2}$}
\author{Edmanuel Torres$^{2}$}
\author{Laurent Karim Béland$^{1, *}$}  

\affiliation{ \vspace{10pt} \textnormal{$^{1}$Department of Mechanical and Materials Engineering, Queen's University, Kingston, ON, Canada  \\ $^{2}$Canadian Nuclear Laboratories, Chalk River Laboratories, Chalk River, Ontario K0J1J0, Canada \vspace{10pt}\\ Correspondence: Karim Zongo (zongo.karim@queensu.ca) or Laurent Karim Béland (laurent.beland@queensu.ca) }}

\begin{abstract} 
\section*{Abstract}

\bfseries{Predicting the properties of multicomponent molten salts using density functional theory (DFT) remains challenging because the spatial and temporal scales required to evaluate transport properties and phase behavior are computationally prohibitive. In this work, we develop a moment tensor potential trained using a a DFT dataset of NaCl, KCl, NaCl–KCl mixtures, and the NaK alloy, enabling large-scale molecular dynamics simulations across wide ranges of temperatures and compositions. We systematically evaluate the effect of D3 dispersion corrections and apply the resulting potential to predict liquid densities, diffusion coefficients, radial distribution functions, heat capacities, thermal conductivities, and the NaCl–KCl phase diagram. The model successfully reproduces many temperature- and composition-dependent trends. However, systematic deviations in several absolute properties persist, highlighting  the importance of experimental validation and calibration. These findings support a hybrid modeling framework in which first-principles-informed machine-learning potentials provide transferable predictive capability and mechanistic insight, while experimental data incorporated during model development or subsequent engineering assessments is necessary to improve quantitative accuracy.
}

\end{abstract}

\maketitle
\section{INTRODUCTION}

Molten salts are promising candidates for high-temperature energy applications because they combine excellent thermal stability with favorable thermophysical and transport properties~\cite{roper2022molten,williams2008evaluation,parker2022thermophysical}. They are widely used, or actively considered, as heat-transfer fluids, coolants, and thermal energy storage media in systems such as concentrated solar power plants and actively considered advanced nuclear reactors~\cite{gonzalez2017review,zinkle2000operating}. In practical applications, these salts are typically deployed as multicomponent mixtures. Therefore, their melting characteristics, density, transport properties, and local structure vary sensitively with composition and temperature. As a result, predictive models must not only capture the properties of pure salts, but also the complex trends that arises across a broad composition space. Developing such models from first principles remains a significant challenge due to the high computational overhead, particularly when accurate absolute property values are required for engineering design~\cite{li2021development}.

Among the molten chloride systems, NaCl--KCl serves as a model case for investigating composition-dependent thermophysical properties and solid--liquid phase equilibria~\cite{coleman1967phase,sergeev2015thermodynamics}. It is chemically simple enough for a systematic first-principles-based modeling, yet sufficiently complex to assess whether an interatomic potential can describe changes in structure, transport, and phase stability across composition. Interactions among Na$^+$, K$^+$, and Cl$^-$ ions can produce nontrivial composition-dependent behavior, including shifts in density, diffusivity, and melting behavior~\cite{bauer2021molten}. Because phase boundaries constrain the range of accessible operating conditions, the ability to reproduce the NaCl--KCl phase diagram provides a stringent test of whether a model can capture the thermodynamic trends needed for more complex molten-salt formulations.

A wide range of experimental and theoretical studies has been devoted to molten salt systems~\cite{janz1979physical,cantor1968physical,li2018experimental,yuan2018experimental,zeidler2022structure,li2021development,ding2023microstructure,bengtson2014first,hazebroucq2005density}. Experimental techniques such as calorimetry, X-ray diffraction, neutron diffraction, and electrochemical measurements provide essential information on structure, thermodynamics, phase stability, and transport properties~\cite{di1997new,bamberger1975experimental,thoma1975phase}. Such measurements are important because they provide the absolute property values needed to validate and calibrate predictive models. However, experimental characterization of molten salts remains difficult because experiments must be carried out at high temperature and highly corrosive environments, often under demanding safety constraints~\cite{williams2006assessment,sridharan2013corrosion}. These issues become more severe in multicomponent systems, where the number of possible compositions grows rapidly and subsequently the exhaustive experimental characterization becomes impractical~\cite{li2021development}.

Atomistic modeling offers a complementary route for studying molten salts under controlled conditions. Molecular dynamics (MD) simulations can resolve structural organization, thermodynamic fluctuations, and transport processes that are difficult to isolate experimentally. DFT provides a description of electronic structure, bonding, and charge redistribution, and is therefore widely used to generate reference data for large-scale simulations~\cite{hohenberg1964inhomogeneous,frandsen2020structure,duemmler2023first}. However, direct DFT simulations are limited to relatively small systems and short times, making it difficult to access long-range correlations, slow transport processes, and phase behavior under realistic thermodynamic conditions. Moreover, DFT predictions can carry systematic errors associated with the chosen exchange--correlation functional, dispersion treatment, and finite simulation cell, so agreement with DFT does not necessarily guarantee agreement with experiment.

To overcome these limitations, various classical interatomic potentials have been developed, including rigid-ion~\cite{sangster1976interionic} and polarizable-ion models~\cite{wilson1993polarization} as well as variable-charge force fields such as ReaxFF~\cite{chenoweth2008reaxff} and COMB~\cite{liang2013classical}. While these approaches enable larger-scale simulations, they rely on predefined functional forms and system-specific parameterizations. Consequently, their transferability across compositions and thermodynamic conditions is often limited, as they often cannot be systematically improved. Extending these models to new chemical systems typically requires extensive reparameterization, which may compromise both transferability and predictive accuracy~\cite{wang2020comparison, van2010development,cheung2005reaxffmgh, liu2011reaxff}.

The broad compositional and thermodynamic landscape of molten salt systems therefore requires modeling approaches which are systematically improvable, transferable, and computationally tractable. Machine-learning interatomic potentials (MLIPs)~\cite{wang2024machine} are emerging as such approach because they can extend first-principles accuracy to the length and time scales required for liquid structure, transport, and phase-equilibrium simulations. Previous neural-network interatomic potentials (NNIPs) have been successfully applied to molten NaCl, yielding predictions of thermophysical properties such as density, heat capacity, diffusion coefficients, thermal conductivity, and melting point in good agreement with experiment~\cite{li2021development}. NNIP-based free-energy calculations have also been used to compute chemical potentials and melting points in LiCl~\cite{gibson2025computing}, while Deep Potential Molecular Dynamics has been applied to several molten chloride systems~\cite{xu2023development,liang2021theoretical,liang2020molecular,guo2022molecular,pan2020dft,liang2022machine,liang2021machine,zhao2021theoretical,pan2021dft} and multicomponent fluoride molten salts~\cite{chahal2022transferable}. In particular, studies of MgCl$_2$--NaCl and MgCl$_2$--KCl eutectic melts show that structural, thermodynamic, and kinetic properties can be captured with MLIP-based simulations~\cite{xu2023development}. Moment Tensor Potentials (MTPs)~\cite{shapeev2016moment} provide an alternative MLIP framework and have been successfully applied to molten NaCl to predict solubility limits, redox potentials, and transport properties while retaining favorable computational efficiency~\cite{sun2024interatomic}.

Despite these advances, two critical issues remain in achieving predictive simulations of multicomponent molten salts. First, most MLIPs have been developed and validated for single-component systems, such as NaCl~\cite{li2021development,sun2024interatomic,tovey2020dft} and LiCl~\cite{gibson2025computing}, or for specific mixtures such as MgCl$_2$--NaCl, MgCl$_2$--KCl~\cite{xu2023development}, LiF--NaF--ZrF$_4$~\cite{chahal2022transferable}, and ZnCl$_2$--NaCl--KCl~\cite{pan2021dft}. The ability of a single potential to describe NaCl, KCl, NaCl--KCl mixtures, and the related NaK alloy across multiple compositions and phases remains less established. Second, even when an MLIP accurately reproduces the first-principles training data, quantitative agreement with experiment may still be limited due systematic errors in the reference data. Dispersion interactions are a clear example. For instance, neglecting van der Waals interactions can produce unphysical volume expansion in molten salts~\cite{liu2014solubility}, but adding an empirical D3 correction is not necessarily a uniform improvement across systems or compositions~\cite{weymuth2014new,kostal2023common}. Its effect on density, structure, and transport may depend on the specific chemistry and thermodynamic state. It is therefore useful to treat dispersion-corrected and uncorrected simulations as related modeling variants, rather than assuming that one is universally more accurate.

In this work, we present a MTP for molten NaCl, KCl, NaCl--KCl mixtures, and the related NaK alloy, trained on DFT data using a small-cell active-learning strategy~\cite{meziere2023accelerating,luo2023set,sun2024interatomic,meng2025small}. We evaluate both the uncorrected MTP and a dispersion-corrected MTP-D3 variant in order to assess the role of long-range dispersion interactions on the predicted properties across different chemistries and compositions. The models are used to evaluate densities, diffusion coefficients, local structural properties, and the NaCl--KCl phase diagram over broad temperature and composition ranges, including compositions not explicitly included in the training dataset. By comparing these predictions with available experimental data, we assess the extent to which the MTP framework to capture both temperature- and composition-dependent trends and quantitatively reproduces absolute property values. This distinction is particularly important for engineering applications, since first-principles-based MLIPs can provide transferable information across composition space, but systematic offsets may require experimental calibration, either during potential training or during downstream property assessment.

\section{COMPUTATIONAL METHODS}

\subsection{Machine-learning interatomic potential}

We employ the MTP formalism~\cite{shapeev2016moment,novikov2021mlip}, which is one of many existing MLIP frameworks for materials modeling~\cite{wang2024machine}. MTPs represent the local atomic environment through systematically constructed tensor descriptors, providing a compact many-body expansion that preserves translational, rotational, and permutational invariance. This framework offers an attractive balance between computational efficiency and predictive accuracy~\cite{zuo2020performance}.

In the MTP formalism, the total energy is expressed as a sum of atomic contributions,
\begin{equation}
E_{\rm Total} = \sum_{i=1}^{N} E_i = \sum_{i=1}^{N} V_{\rm local}(r_i),
\end{equation}
where $r_i=(r_{ij})_{j=1}^{n_i}$ denotes the local environment of atom $i$, and $r_{ij}=x_j-x_i$ is the position vector of neighboring atom $j$ relative to atom $i$. The local energy $V_{\rm local}$ is evaluated using atoms within a cutoff radius $R_c$, set to 7.5~\AA\ in this work, and expanded in terms of basis functions $B_\beta$:
\begin{equation}
V_{\rm local} = \sum_\beta c_\beta B_\beta .
\end{equation}

The basis functions $B_\beta$ are constructed by contracting moment tensor descriptors that encode the radial and angular distributions of neighboring atoms:
\begin{multline}
M_{\mu,\nu}(r_{i}) = 
\sum_j f_\mu(|r_{ij}|, \tau_i, \tau_j) 
\underbrace{r_{ij} \otimes \cdots \otimes r_{ij}}_{\nu \text{ times}},
\label{eq:muv}
\end{multline}
where $\tau_i$ and $\tau_j$ denote the chemical species of atoms $i$ and $j$, respectively. The radial functions $f_\mu$ are expanded in radial basis functions $Q^{(\alpha)}$ as
\begin{equation}
f_\mu(|r_{ij}|, \tau_i, \tau_j) = 
\sum_\alpha c_{\mu,\tau_i,\tau_j}^{(\alpha)} Q^{(\alpha)}(|r_{ij}|).
\end{equation}
The functions $Q^{(\alpha)}(r)$ are defined on the interval $r \in [R_{\min}, R_c]$ and are constructed to ensure $C^1$ continuity at the cutoff. The model parameters,
$\Theta = \{c_\beta, c_{\mu,\tau_i,\tau_j}^{(\alpha)}\}$, are obtained by fitting to reference energies, forces, and stresses.

The complexity of the MTP is controlled by the maximum level of the retained basis functions. Each moment $M_{\mu,\nu}$ is assigned a level,
\begin{equation}
\text{levM}_{\mu,\nu} = 2 + 4\mu + \nu,
\end{equation}
and the level of a basis function $B_\beta$ formed by contractions of multiple moments is
\begin{equation}
\text{levB}_\beta = \sum_i \left( 2 + 4\mu_i + \nu_i \right).
\end{equation}
Only basis functions satisfying
\begin{equation}
\text{levB}_\beta \leq \text{lev}_{\max}
\end{equation}
are retained. The value of $\text{lev}_{\max}$ therefore defines the model complexity. In this work, we use a level-22 MTP. Unless otherwise specified, this MTP refers to the local machine-learning potential trained on DFT reference data; dispersion-corrected simulations are treated separately as an MTP-D3 variant.

\subsection{Ab initio calculations}

The reference dataset for training and validating the interatomic potentials was generated from DFT calculations carried out with the Quantum ESPRESSO package~\cite{giannozzi2009quantum,giannozzi2017advanced}. Ultrasoft pseudopotentials~\cite{kresse1999ultrasoft} were employed with the generalized gradient approximation~\cite{perdew1996generalized}, using the Perdew--Burke--Ernzerhof functional~\cite{ernzerhof1999assessment} to describe the exchange--correlation energy. Unless otherwise specified, dispersion corrections were not included in the DFT reference calculations; their effect is assessed separately in molecular dynamics simulations. The kinetic energy cutoff for the plane-wave basis was set to 900~eV, based on convergence tests for total energies, forces, and stress tensors across representative configurations. Because the database includes configurations with different numbers of atoms, cell volumes, densities, and thermodynamic states, Monkhorst--Pack k-point meshes~\cite{monkhorst1976special} were chosen according to cell size to maintain consistent reciprocal-space sampling. These DFT calculations define the reference potential-energy surface learned by the MTP, and any systematic errors associated with the chosen functional or missing long-range interactions may therefore be inherited by the trained model.

\subsection{Molecular Dynamics Simulations and Active-Learning-Based Small-Cell Training}

All MD simulations were performed using the LAMMPS package~\cite{thompson2022lammps} with a time step of 1~fs. Depending on the property being evaluated, simulations were carried out in the NVE, NVT, NPT, or NPH ensemble. Temperature and pressure were controlled using the Nosé--Hoover thermostat~\cite{evans1985nose} and the Parrinello--Rahman barostat~\cite{parrinello1981polymorphic}, respectively, with damping parameters of 100~fs and 1000~fs. Unless otherwise stated, simulations used the DFT-trained MTP described above; simulations including the D3 dispersion correction are indicated as MTP-D3. Additional simulation details for each property calculation are provided in the main text and Supporting Information.

We applied active learning to construct the training dataset efficiently~\cite{podryabinkin2017active}. Within the MTP framework~\cite{podryabinkin2017active,podryabinkin2023mlip}, configurations generated during MD are evaluated on-the-fly using an extrapolation grade, $\gamma$, which measures how far a configuration lies outside the domain spanned by the current training set. Configurations with $2 < \gamma < 10$ were selected for DFT calculations and added to the database, whereas simulations were terminated when $\gamma > 10$ as the potential was considered unreliable in such case. The training dataset was then updated, the potential retrained, and the active dataset reconstructed in an iterative loop. This procedure enables efficient exploration of relevant configurational space while limiting the number of reference DFT calculations. A full theoretical description of the active-learning scheme is provided elsewhere~\cite{podryabinkin2019accelerating,podryabinkin2017active,podryabinkin2023mlip}.

Although active learning reduces the cost of exploring configurational space, the efficiency of MLIP development also depends strongly on the size and diversity of the reference database. Training data generated from DFT is expensive, particularly for molten salts where relevant configurations span multiple compositions, densities, temperatures, and phases. Expanding the dataset can improve robustness and transferability, but also increases the cost of data generation, training, and validation. To address this challenge, we use a small-cell training strategy, in which the potential is trained on configurations containing relatively few atoms~\cite{pickard2022ephemeral,meziere2023accelerating,luo2023set,meng2025small}. Small-cell datasets can be generated using static structure-search approaches, such as AIRSS~\cite{pickard2022ephemeral}, USPEX~\cite{oganov2006crystal,oganov2011evolutionary,lyakhov2013new}, and RANDSPG~\cite{avery2017randspg}, or dynamically through active-learning MD simulations~\cite{meziere2023accelerating,luo2023set,sun2024interatomic,meng2025small}. Here, we combine small-cell training with active learning to construct a transferable MTP for NaCl, KCl, NaCl--KCl mixtures, and NaK, with the training dataset biased toward molten-state configurations.

The initial database was constructed from primitive cells of Na, K, NaCl, and KCl obtained from crystallographic repositories~\cite{ceder2010materials,downs2003american}, with additional configurations generated by applying volumetric strains. A preliminary MTP was trained on this dataset together with reference data from Ref.~\cite{sun2024interatomic}. The database was then expanded using active learning with selection and termination thresholds of $\gamma_{\mathrm{select}}=2$ and $\gamma_{\mathrm{break}}=10$, respectively. Active learning was performed sequentially for systems containing $N=2$ to $N=36$ atoms ($N=2,4,8,12,16,24,34,36$). At each system size, parallel MD simulations were carried out over temperatures spanning solid and liquid regimes. Configurations identified as extrapolative were filtered for redundancy, recomputed using DFT, added to the training set, and used to retrain the MTP. This procedure was repeated until no configurations exceeded $\gamma_{\mathrm{select}}$.

The dataset was subsequently extended to binary systems. For NaK, small cells with multiple compositions, including the eutectic composition of 78\% K and 22\% Na, were sampled using active-learning MD in the liquid phase. A similar procedure was used for NaCl--KCl mixtures, including equimolar compositions. In total 3721 configurations were generated. After curation, 1277 configurations were retained for training and 200 for validation across Na, K, NaCl, KCl, NaK, and NaCl--KCl systems. A level-22 MTP with 935 parameters was used for the final model. The curated database is strongly biased toward liquid-phase configurations ($\sim$98\%), consistent with the focus of this work on molten-state properties, while still retaining representative solid-state configurations needed for phase-equilibrium calculations. Additional details are provided in the Supporting Information.

\begin{figure}
\centering
    \includegraphics[width=6.8cm]{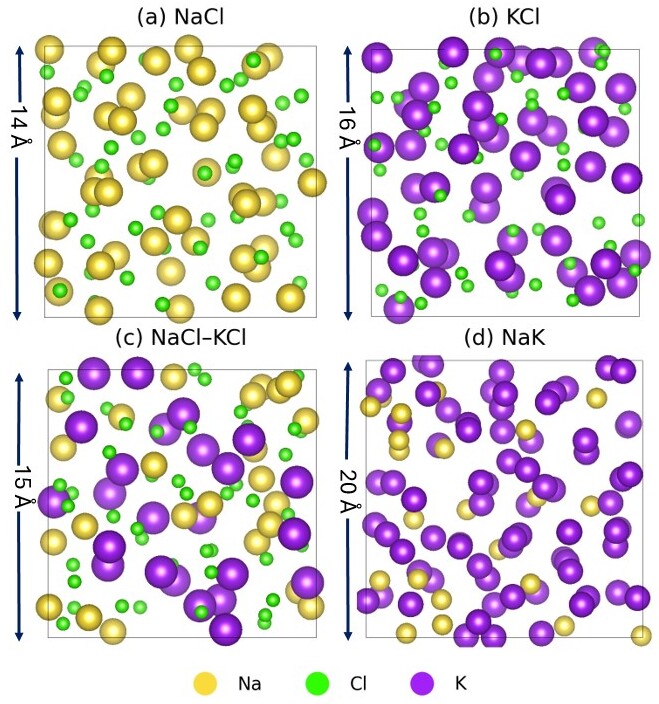}
    \caption{Atomic structures of (a) molten NaCl at 1150~K, (b) molten KCl at 1100~K, (c) molten equimolar NaCl--KCl at 1150~K, and (d) liquid metallic NaK at 300~K. Each system contains 96 atoms and is modeled using a cubic simulation cell.}
    \label{fig:molten-salt}
\end{figure}

\subsection{Dispersion correction D3}

The DFT reference data used to train the MTP was generated without dispersion corrections. Therefore, we considered an additional MTP-D3 variant in which a D3 dispersion correction~\cite{grimme2010consistent,grimme2011effect,qamar2023atomic} is added during molecular dynamics simulations. This correction was introduced to assess the sensitivity of predicted properties to long-range interactions that are not explicitly included in the DFT reference data or in the MTP. As reported in Ref.~\cite{liu2014solubility}, neglecting van der Waals interactions can lead to unphysical volume expansion in molten salts, although the effect of an empirical dispersion correction need not be uniformly beneficial across different compositions or properties. A convergence test was performed for molten NaCl and KCl at 1150~K to determine the parameters for the D3 correction. Zero-damping and Becke--Johnson (BJ) damping schemes were tested, and NPT simulations were carried out while monitoring density and total energy. The MTP cutoff radius was 7.5~\AA, while the D3 cutoff was varied from 5 to 25~\AA. Both energy and density were converged for D3 cutoffs above 20~\AA. Thus, we employ a 20~\AA\,cutoff with the BJ damping when the MTP-D3 variant is used.

\subsection{Simulation of melting point and phase diagram}

\begin{figure}
\centering
    \includegraphics[width=8.5cm]{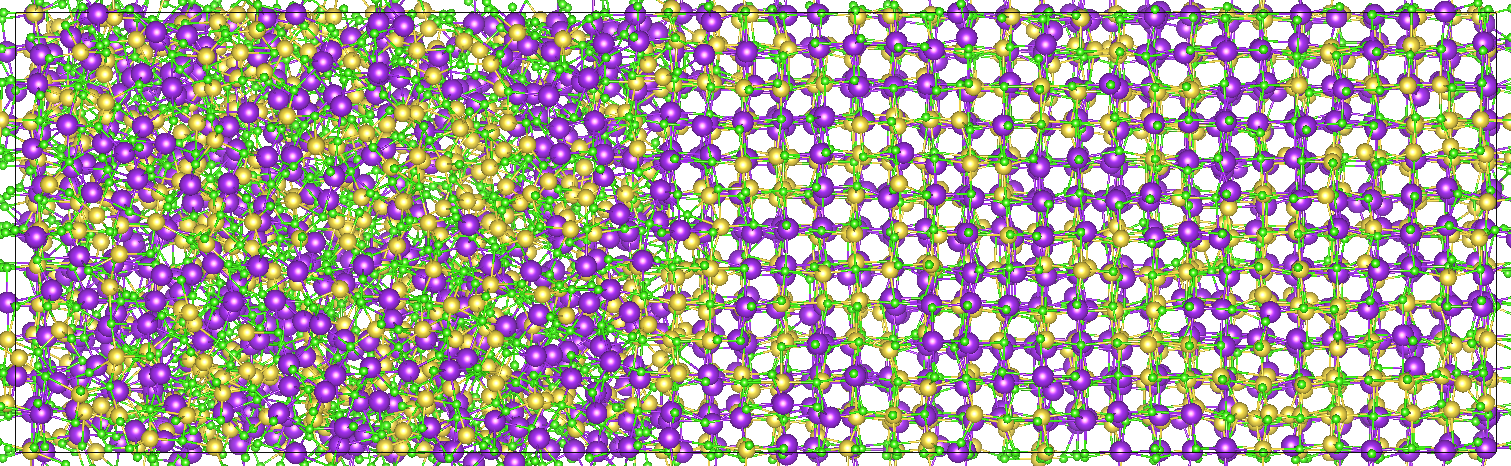}
    \caption{Illustration of the solid--liquid coexistence model used to determine melting temperatures and phase boundaries in the NaCl--KCl system. Each simulation cell contains 5760 atoms and is constructed in a tetragonal box elongated along the $z$-direction to accommodate the solid--liquid interface. The configuration shown corresponds to a NaCl--KCl mixture containing 50 mol\% NaCl and is representative of the simulation setup used to estimate solidus and liquidus temperatures across the composition range.}
    \label{fig:CMI}
\end{figure}

As a stringent test of composition-dependent phase behavior, the NaCl--KCl phase diagram was simulated over the composition range of 10--90 mol\% NaCl, with pure NaCl and pure KCl serving as the compositional endpoints. All simulations were performed at zero pressure to estimate solid--liquid equilibrium temperatures. The moving-interface method~\cite{zhang2012comparison} was employed for both the pure salts and intermediate compositions. Simulations were carried out in the NPT ensemble using systems containing a solid--liquid interface (Fig.~\ref{fig:CMI}) at a series of temperatures near the expected transition region. Production simulations were performed for 1~ns, and five independent configurations were generated for each composition. For intermediate compositions, independent configurations were constructed by randomly substituting Na ions with K ions using different random seeds, thereby sampling multiple cation arrangements at fixed composition.

For the pure end members, the solid--liquid interface migrated toward either the solid or liquid phase depending on whether the simulation temperature was below or above the melting temperature. The total energy was monitored as a function of time, and the rate of energy change was obtained from a linear fit. The melting temperature was identified as the temperature at which the energy-change rate approached zero, corresponding to a stationary interface and thermodynamic coexistence between the solid and liquid phases.

For intermediate NaCl--KCl compositions, the same moving-interface procedure was used to estimate the solid--liquid transition interval. Simulations were conducted at multiple temperatures, and the evolution of both the interface and total energy was monitored throughout each trajectory. Rather than assigning a single sharp melting temperature, the coexistence region was bracketed by identifying the highest temperature at which the system remained solid and the lowest temperature at which the system became fully liquid within the simulation time. The difference between these bounds defines the estimated coexistence width for each composition. This procedure provides a composition-dependent phase-boundary estimate that can be compared with experiment to evaluate both trend fidelity and systematic offsets in the predicted phase diagram.

\subsection{Free energy calculation}

The miscibility gap provides a thermodynamic test of composition-dependent mixing energetics and phase stability in multicomponent systems. Here, we use thermodynamic integration (TI)~\cite{freitas2016nonequilibrium} to compute Gibbs free energies of mixing for solid NaCl--KCl mixtures and to estimate the corresponding miscibility gap. This analysis is used not only to compare with available phase-boundary information, but also to assess whether the MTP captures the thermodynamic trends associated with mixing, as distinct from possible systematic offsets in absolute free energies.

Models of NaCl--KCl mixtures with varying compositions were generated using a two-step procedure. First, the endpoint compounds, NaCl and KCl, were relaxed using the MTP to determine their equilibrium lattice parameters. For a mixture with NaCl mole fraction $x$, the initial lattice parameter was estimated using Vegard's law~\cite{denton1991vegard},
\begin{equation}
a(x)=x\,a_{\rm NaCl}+(1-x)\,a_{\rm KCl},
\end{equation}
where $a_{\rm NaCl}$ and $a_{\rm KCl}$ are the relaxed lattice parameters of the pure compounds. A supercell was then constructed using the interpolated lattice parameter, and a fraction $1-x$ of the Na sites was randomly substituted by K atoms to obtain the desired NaCl--KCl composition.

For each composition, molecular dynamics simulations were performed from 300 to 950~K in increments of 50~K. Each simulation consisted of energy minimization, gradual heating in the NVT ensemble, and equilibration under pressure-controlled dynamics for 100~ps. Production runs were then performed for 200~ps, during which thermodynamic and structural quantities, including total energy, density, lattice parameters, and species-resolved mean-square displacements (MSDs), were collected. The MSDs were used to determine effective harmonic spring constants for the subsequent TI calculations, and statistical uncertainties were estimated using block averaging.

Thermodynamic integration was performed at 15 temperatures between 300 and 1000~K using the Frenkel--Ladd method~\cite{freitas2016nonequilibrium} with the precomputed temperature-dependent harmonic spring constants. Initial configurations were taken from equilibrated MSD production trajectories. The TI simulations used a timestep of 1~fs, with equilibration and integration periods of 5000 steps each. Forward and backward switching trajectories were performed at each temperature, and the derivative of the energy with respect to the coupling parameter was recorded along the integration path. Absolute free energies were obtained by combining the TI contribution with the Einstein-crystal reference free energy, including the center-of-mass correction.

The resulting free energies were used to evaluate the Gibbs free energy of mixing as a function of composition and temperature. This provides a thermodynamic measure of the stability of homogeneous NaCl--KCl solid solutions relative to phase-separated reference states. In particular, the curvature and temperature dependence of the free-energy curves were used to assess the tendency toward mixing or demixing. For a binary system with composition $x$, the Gibbs free energy of mixing relative to the unmixed end-member reference state is

\begin{equation}
\begin{aligned}
\Delta G_{\mathrm{mix}}(x,T)
=&\; G_{\mathrm{hom}}(x,T)
- \left[xG_A(T) + (1-x)G_B(T)\right] \\
&- T\Delta S_{\mathrm{mix}} ,
\end{aligned}
\end{equation}

where $G_{\mathrm{hom}}(x,T)$ is the Gibbs free energy of the homogeneous mixture, $G_A(T)$ and $G_B(T)$ are the Gibbs free energies of the pure end members at the same temperature and pressure, and $\Delta S_{\mathrm{mix}}$ is the configurational entropy of mixing. Assuming ideal configurational entropy,

\begin{equation}
\Delta S_{\mathrm{mix}}
=
-R\left[x\ln x + (1-x)\ln(1-x)\right],
\end{equation}

where $R$ is the gas constant. For the NaCl--KCl system, with $x$ denoting the mole fraction of NaCl, this gives

\begin{equation}
\begin{aligned}
\Delta G_{\mathrm{mix}}(x,T)
=&\; G_{\mathrm{NaCl\text{-}KCl}}(x,T) \\
&- \left[
xG_{\mathrm{NaCl}}(T)
+(1-x)G_{\mathrm{KCl}}(T)
\right] \\
&+ RT\left[
x\ln x
+(1-x)\ln(1-x)
\right].
\end{aligned}
\end{equation}

Because the logarithmic term is negative for $0<x<1$, this ideal entropic contribution stabilizes the homogeneous mixture, whereas the energetic contribution from the computed free energies may favor demixing. The competition between these terms determines whether mixing is thermodynamically favorable at a given composition and temperature.

To estimate the temperature at which a homogeneous mixture becomes stable relative to the unmixed end-member reference state, $\Delta G_{\mathrm{mix}}(x,T)$ was evaluated at fixed composition as a function of temperature. For the $x=0.10$ composition, the condition $\Delta G_{\mathrm{mix}}=0$ was reached within the simulated temperature range, and the corresponding transition temperature was obtained by linear interpolation. For the remaining compositions, $\Delta G_{\mathrm{mix}}$ did not reach zero over the simulated temperature range. In these cases, the transition temperature was estimated by fitting the approximately linear region of the $\Delta G_{\mathrm{mix}}(T)$ curve and extrapolating to $\Delta G_{\mathrm{mix}}=0$. The fitting interval was selected separately for each composition based on the temperature range over which the data showed approximately linear behavior. These extrapolated temperatures should therefore be interpreted as estimates of the mixing stability boundary within the assumptions of the ideal configurational entropy model and the chosen end-member reference state.

\subsection{Comparison of computational cost and performance: MTP vs MTP--D3}

\begin{figure}
\centering
    \includegraphics[width=8.5cm]{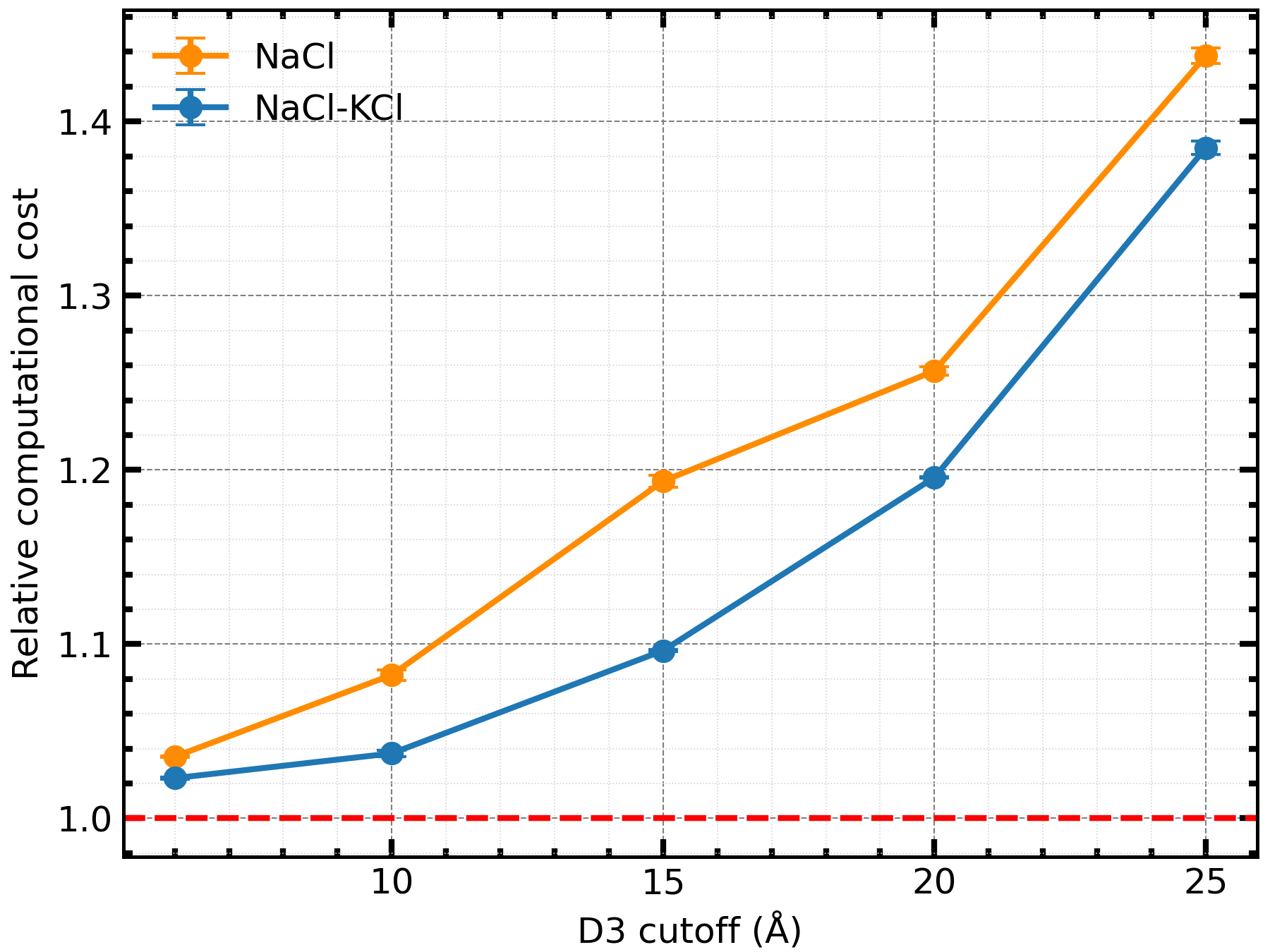}
    \caption{Computational performance comparison for molten NaCl and equimolar NaCl--KCl systems. The reported cost corresponds to the wall-clock time per MD step and is normalized with respect to MTP-only simulations performed under identical thermodynamic and parallelization conditions. Measurements were carried out on CPU using simulation cells containing 16,000 atoms.}
    \label{fig:cost}
\end{figure}

The computational performance of the baseline MTP and the MTP-D3 variant was evaluated to quantify the overhead associated with adding long-range dispersion corrections in large-scale molten-salt simulations. Reported timings correspond to full molecular dynamics production runs, including force evaluation, neighbor-list construction, and MPI communication overhead, thereby providing an application-level measure of computational cost. All benchmarks were performed on systems containing 16,000 atoms using 64 MPI processes. Performance is reported as wall-clock time per MD step, and multiple independent runs with different initial velocity seeds were used to reduce statistical fluctuations. The MTP-D3 simulations employed BJ damping and a converged 20~\AA\ D3 cutoff. Production runs consisted of 300,000 MD steps and are summarized in Fig.~\ref{fig:cost}.

The baseline MTP exhibits a computational cost of $\sim 10^{-7}$ core\,h\,atom$^{-1}$\,MD step$^{-1}$, consistent with previous work~\cite{sun2024interatomic} and approximately one order of magnitude lower than reported neural-network interatomic potentials for molten salts~\cite{li2021development}. Adding the D3 correction increases the cost by $\sim$25\% for molten NaCl and by $\sim$20\% for equimolar NaCl--KCl under the benchmark conditions considered here. Thus, the long-range dispersion correction introduces a modest computational overhead while preserving the favorable efficiency of the MTP framework. This benchmark quantifies the cost of the MTP-D3 variant, but does not imply that the dispersion-corrected model is uniformly more accurate across compositions or properties.

\section{Results and discussion}

\subsection{Bulk properties}

The MTP predicts bulk moduli of 30.7 and 16.6~GPa for NaCl and KCl, respectively, compared with experimental values of approximately 24 and 17~GPa~\cite{froyen1986structural}. The predicted bulk modulus of KCl is therefore in good agreement with experiment, whereas that of NaCl is noticeably overestimated. This discrepancy illustrates that, even when a MLIP accurately reproduces its first-principles reference data, absolute properties may still inherit systematic errors from the underlying DFT description.

\begin{figure}
\centering
    \includegraphics[width=8.5cm]{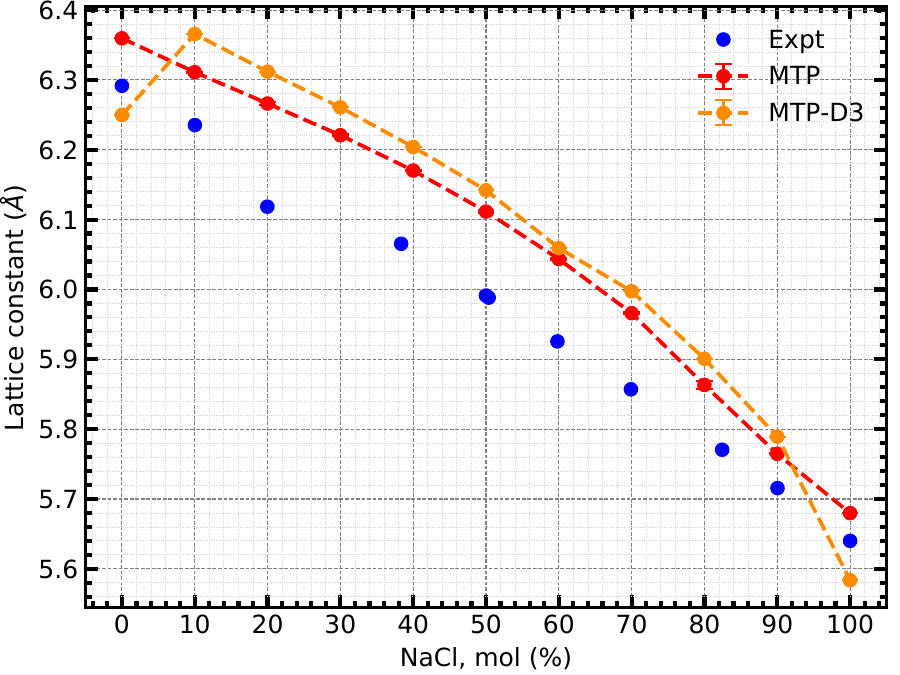}
    \caption{Lattice parameters of NaCl--KCl mixtures as a function of NaCl composition, predicted using the baseline MTP and the MTP-D3 variant and compared with experimental data. The lattice parameters at the end-member compositions were obtained by energy minimization, whereas those at intermediate compositions were determined by relaxation to zero pressure at 0~K.}
    \label{fig:latt-param}
\end{figure}

Fig.~\ref{fig:latt-param} presents the lattice parameters of NaCl, KCl, and NaCl--KCl mixtures as a function of composition, comparing predictions from the baseline MTP and the MTP-D3 variant with experimental data~\cite{barrett1954studies,froyen1986structural}. The baseline MTP gives better agreement with experiment than MTP-D3 across the compositions considered. Thus, for these crystalline lattice parameters, adding the D3 correction does not improve the prediction and instead introduces an additional systematic contraction. This behavior emphasizes that the effect of dispersion corrections is property- and composition-dependent, rather than uniformly beneficial.

\subsection{MTP predictions on the liquid-phase atomic structures}

\begin{figure}
\centering
    \includegraphics[width=6.8cm]{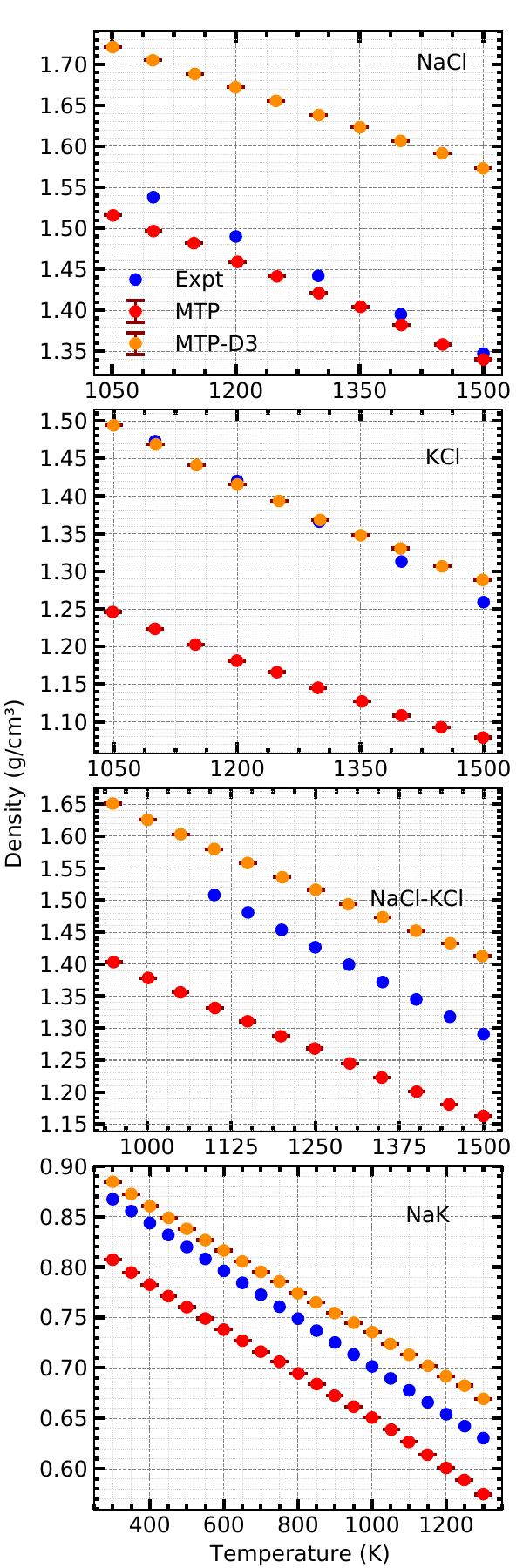}
    \caption{Temperature-dependent densities of molten NaCl, KCl, equimolar NaCl--KCl, and liquid NaK, predicted using the baseline MTP and the MTP-D3 variant and compared with experimental measurements. Error bars are negligible on the scale of the plot. Densities were obtained from simulations containing 1728 atoms, initialized in cubic simulation cells.}
    \label{fig:dens}
\end{figure}

To assess liquid-phase thermophysical properties, the densities of molten NaCl, KCl, equimolar NaCl--KCl, and the liquid NaK alloy were calculated over a range of temperatures and compared with available experimental data, as shown in Fig.~\ref{fig:dens}. Three sets of results are presented: predictions from the baseline MTP, predictions from the MTP-D3 variant, and experimental reference values~\cite{kirshenbaum1962density,janz1979physical,mozgovoi2003density}. Because densities were evaluated over broad temperature ranges, we compare the maximum deviation from experiment for each system.

The effect of the D3 correction is strongly system dependent. For molten KCl, the maximum deviation from experiment decreases from 15.74\% with the baseline MTP to 2.35\% with MTP-D3, indicating a substantial improvement. For equimolar NaCl--KCl, however, the maximum deviation changes only slightly, from 9.90\% to 9.47\%. For liquid NaK, the deviation decreases from 8.79\% to 6.25\%. In contrast, molten NaCl shows the opposite behavior: the maximum deviation increases from 1.45\% with the baseline MTP to 16.78\% with MTP-D3. Thus, adding D3 does not provide a uniform improvement in density predictions. Instead, it introduces composition-dependent shifts in the absolute density, while the temperature-dependent trends remain broadly consistent. This behavior supports treating MTP and MTP-D3 as related modeling variants rather than assuming that the dispersion-corrected model is universally more accurate.

\begin{figure*}
\centering
    \includegraphics[width=18cm]{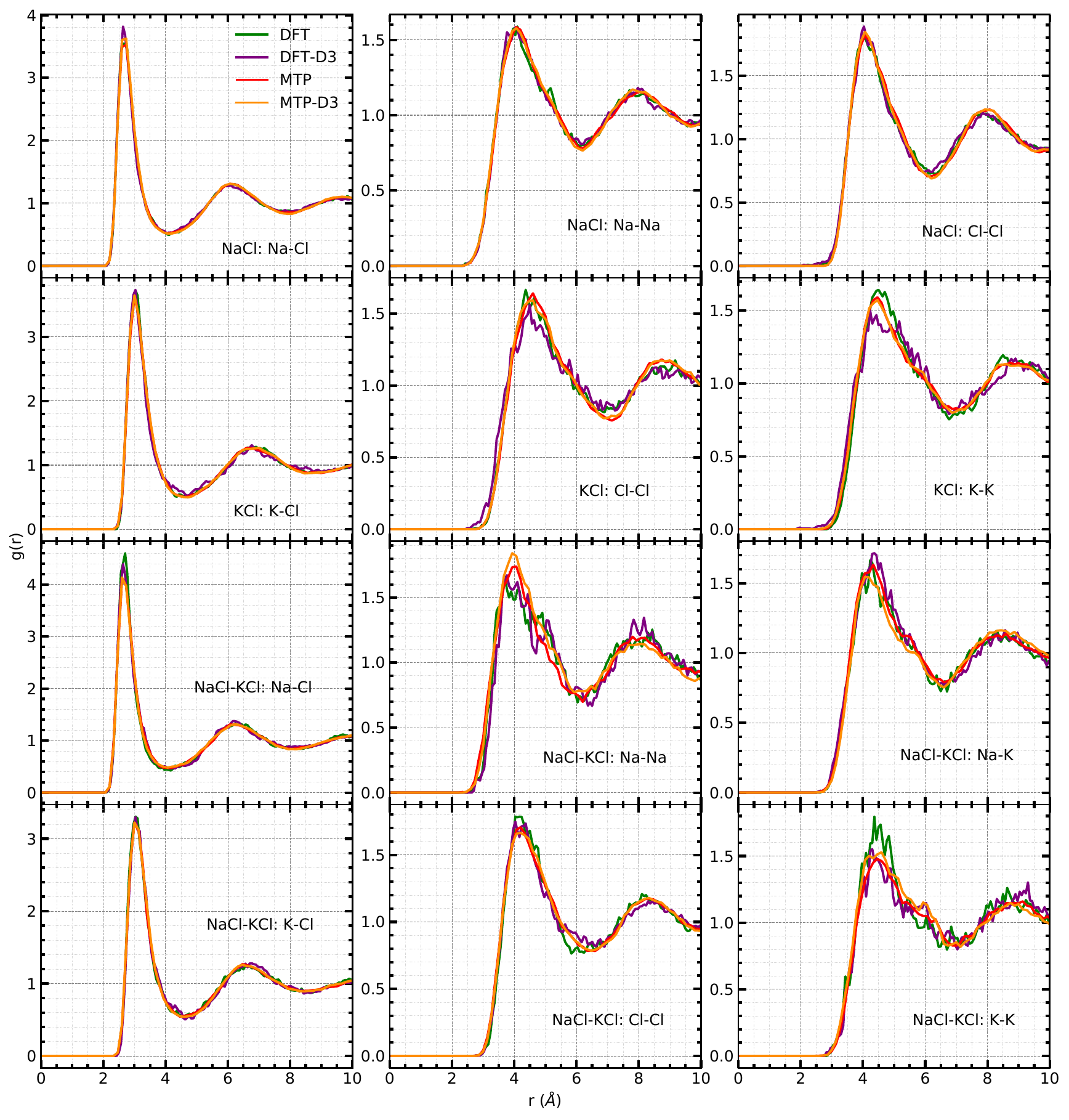}
    \caption{Partial radial distribution functions of molten NaCl, KCl, and equimolar NaCl--KCl from 96-atom simulations. MTP and MTP-D3 predictions are benchmarked against DFT and DFT-D3 calculations. Simulation cell densities were fixed to experimental values at 1150~K for NaCl and 1100~K for KCl and NaCl--KCl, with simulations carried out at the corresponding temperatures.}
    \label{fig:RDF}
\end{figure*}

To further characterize the liquid structure, partial RDFs were computed for molten NaCl, KCl, and equimolar NaCl--KCl. Four datasets were compared: DFT, DFT-D3, MTP, and MTP-D3, as shown in Fig.~\ref{fig:RDF}. For all three systems, the MTP and MTP-D3 RDFs closely reproduce the corresponding DFT-based liquid structures, indicating that the local ordering learned by the MTP is consistent with the first-principles reference data.

Only minor differences are observed among the four datasets. This limited sensitivity should be interpreted in light of the simulation protocol: the RDF calculations were performed at fixed simulation cell densities corresponding to experimental values. Fixing the volume constrains the average bonding environment and suppresses the density changes observed in NPT simulations. The RDF results therefore show that both MTP variants reproduce local liquid structure at fixed density, while the density calculations in Fig.~\ref{fig:dens} demonstrate that absolute volume predictions remain sensitive to the treatment of dispersion.

\subsection{Thermophysical and transport properties}

Thermophysical and transport properties are central to the engineering use of molten salts. Here, we evaluate thermal conductivity, constant-pressure heat capacity, and self-diffusion coefficients using the MTP-D3 variant. This choice provides a consistent basis for the property calculations reported in this section, but should not be interpreted as implying that the D3-corrected model is uniformly more accurate for all properties or compositions, as discussed above for the density and lattice-parameter predictions.

Self-diffusion coefficients were computed from simulations in the canonical ensemble. The MSD was evaluated as a function of time, and diffusion coefficients were obtained from the slope of the linear diffusive regime using the Einstein relation. The isobaric heat capacity was computed from enthalpy fluctuations in the NPT ensemble, with uncertainties estimated by block averaging over multiple independent trajectories. Thermal conductivity was calculated after NPT equilibration using the Müller--Plathe reverse non-equilibrium molecular dynamics method, in which a heat flux is imposed through periodic velocity exchanges and the resulting temperature gradient is measured. Detailed simulation protocols and analysis procedures are provided in the Supporting Information.

\begin{figure}
\centering
    \includegraphics[width=6.8cm]{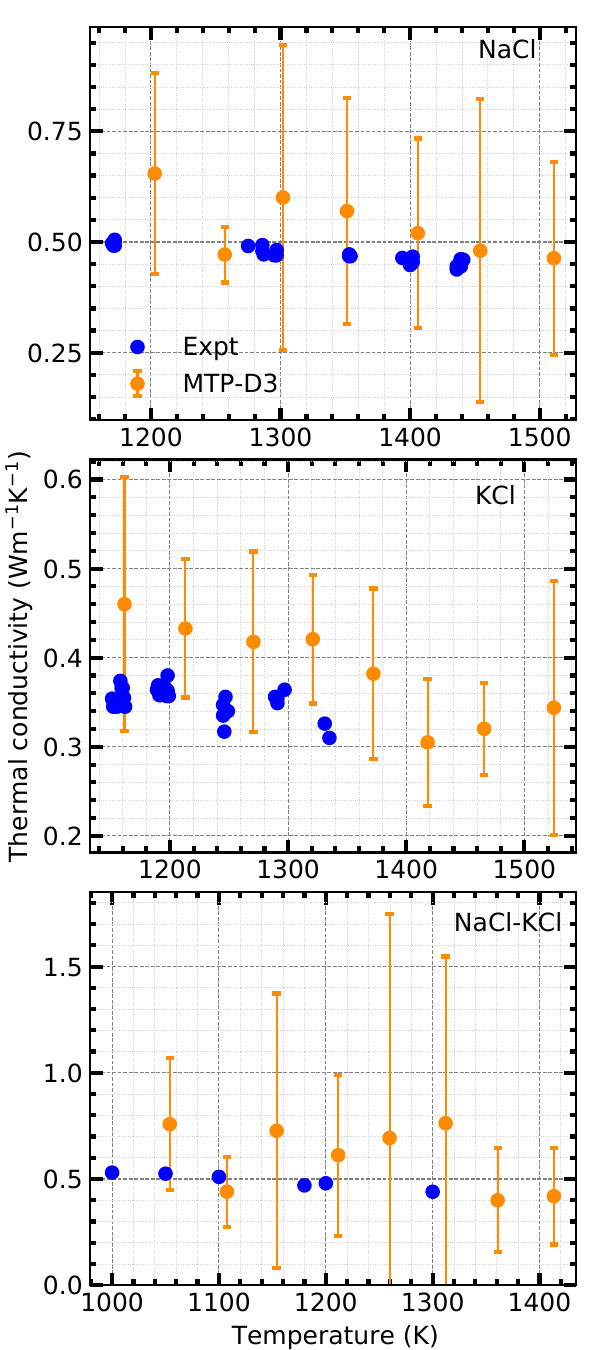}
    \caption{Thermal conductivity of molten NaCl, KCl, and equimolar NaCl--KCl predicted using the MTP-D3 variant and compared with experimental measurements. Error bars denote statistical uncertainties from the simulations.}
    \label{fig:therm-cond}
\end{figure}

Fig.~\ref{fig:therm-cond} presents the predicted thermal conductivity of molten NaCl, KCl, and the equimolar NaCl--KCl mixture, together with available experimental data~\cite{nagasaka1992experimental,lonergan2023thermodynamic,takase2012thermal}. The MTP-D3 predictions are broadly consistent with the experimental ranges for all three systems, although the scatter in the experimental data and the statistical uncertainties in the simulations limit the precision of this comparison. These results indicate that the model captures the approximate magnitude and temperature dependence of thermal conductivity for the systems considered.

\begin{figure}
\centering
    \includegraphics[width=7cm]{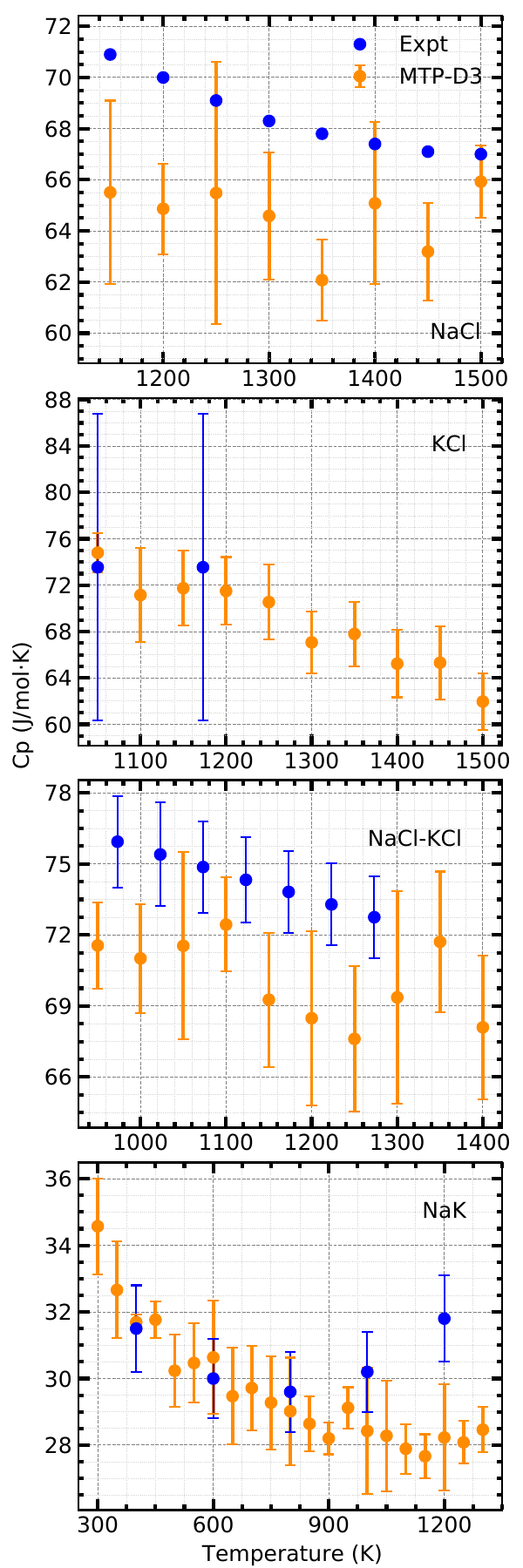}
    \caption{Constant-pressure heat capacity of molten NaCl, KCl, equimolar NaCl--KCl, and liquid NaK as a function of temperature. MTP-D3 predictions with statistical uncertainties are compared with available experimental data.}
    \label{fig:cp-heat-cap}
\end{figure}

Using the same MTP-D3 variant, we also evaluated the constant-pressure heat capacity of molten NaCl, KCl, equimolar NaCl--KCl, and the liquid NaK alloy. The results are shown in Fig.~\ref{fig:cp-heat-cap} and compared with available experimental measurements~\cite{chase1998nist,dubinin2025heat}. The predicted heat capacities are in reasonable agreement with experiment across the investigated temperature range, suggesting that the model captures the main thermal response of the liquid systems.

\begin{figure}
\centering
    \includegraphics[width=6.8cm]{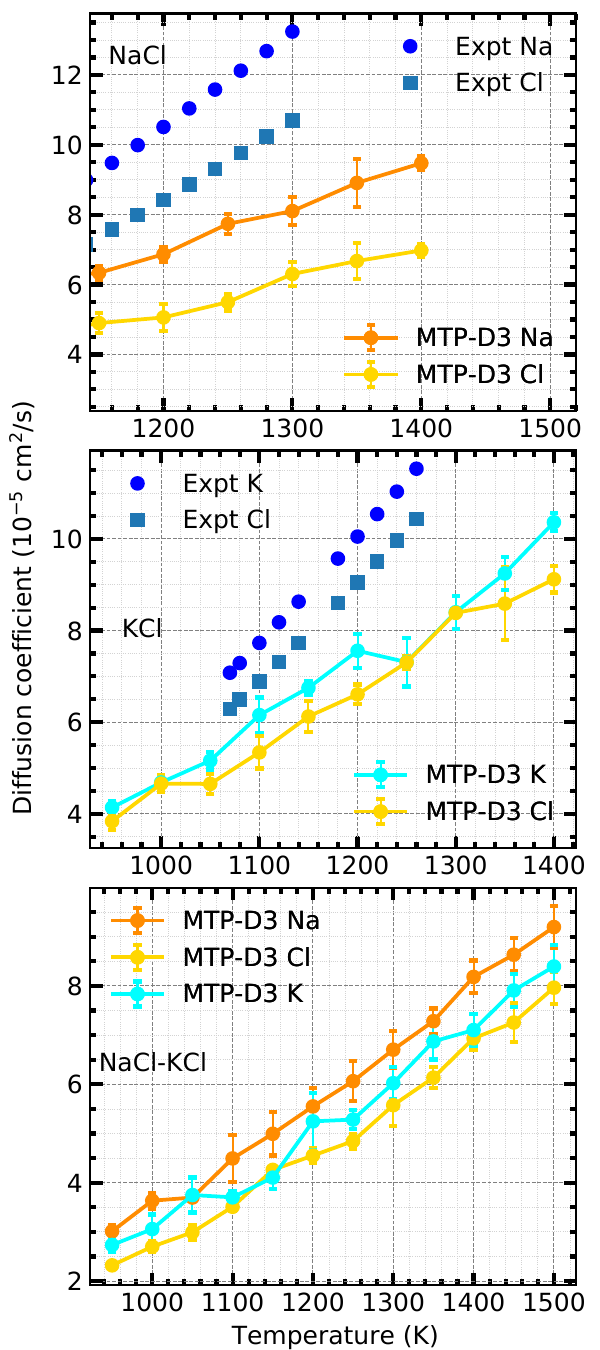}
    \caption{Temperature-dependent self-diffusion coefficients of Na$^+$, K$^+$, and Cl$^-$ in molten NaCl, KCl, and equimolar NaCl--KCl. MTP-D3 predictions, including statistical uncertainties, are compared with available experimental data.}
    \label{fig:dif}
\end{figure}

Finally, self-diffusion coefficients were computed for molten NaCl, KCl, and equimolar NaCl--KCl. As shown in Fig.~\ref{fig:dif}, the predicted diffusion coefficients increase monotonically with temperature for Na$^+$, K$^+$, and Cl$^-$, consistent with experimental trends~\cite{janz1979physical}. However, the absolute values are systematically underestimated across the systems considered. This discrepancy appears to be systematic rather than system-specific, and may reflect limitations inherited from the underlying DFT reference data, the dispersion treatment, or the simulated liquid structure and density. The diffusion results therefore reinforce the broader conclusion that the MTP framework can capture useful temperature-dependent trends, while absolute transport coefficients may require experimental anchoring or calibration for engineering use.

\begin{figure}
\centering
    \includegraphics[width=8.5cm]{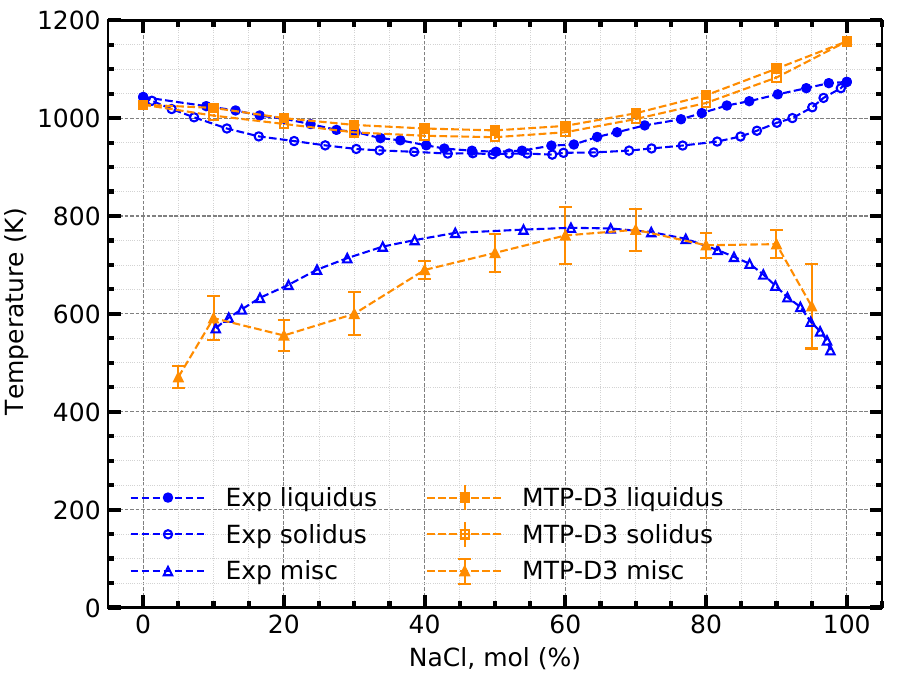}
    \caption{Temperature--composition phase diagram of NaCl--KCl mixtures, including predicted solidus and liquidus boundaries and the estimated miscibility gap. MTP-D3 predictions are compared with experimental data for NaCl compositions from 10 to 90 mol\%, with pure NaCl and KCl shown as reference endpoints.}
    \label{fig:misc-gap}
\end{figure}

\subsection{Melting points and phase diagram}

The predicted melting temperatures of the end-member salts obtained using the MTP-D3 variant are $1027.4 \pm 4.2~\mathrm{K}$ for KCl and $1156.9 \pm 4.0~\mathrm{K}$ for NaCl. These values can be compared with experimental melting temperatures of 1043~K and 1073~K, corresponding to deviations of 1.5\% and 7.8\%, respectively. The KCl melting point is therefore reproduced closely, whereas the NaCl melting point exhibits a noticeable upward shift. This difference is consistent with the broader observation that first-principles-derived potentials can capture temperature- and composition-dependent trends while retaining systematic offsets in absolute thermodynamic quantities. Such offsets may arise from the underlying DFT reference data, finite-size effects, the treatment of dispersion interactions, or the details of the solid--liquid coexistence procedure.

The predicted enthalpies of fusion, $\Delta H_{\mathrm{m}}$, evaluated at the corresponding melting temperatures are $24.30 \pm 0.03~\mathrm{kJ\,mol^{-1}}$ for NaCl and $27.02 \pm 0.03~\mathrm{kJ\,mol^{-1}}$ for KCl. These values compare with experimental measurements of $28.16 \pm 0.14~\mathrm{kJ\,mol^{-1}}$ for NaCl and $26.53 \pm 0.27~\mathrm{kJ\,mol^{-1}}$ for KCl. The KCl enthalpy of fusion is in close agreement with experiment, while the NaCl value is underestimated. Thus, as for the melting temperatures, the end-member thermodynamic properties show good qualitative consistency with experiment but also reveal system-dependent quantitative deviations.

The calculated solidus and liquidus boundaries for NaCl--KCl compositions between 10 and 90~mol\% NaCl are presented in Fig.~\ref{fig:misc-gap}, together with the estimated miscibility limits and experimental phase-diagram data. The MTP-D3 variant reproduces the main topology of the NaCl--KCl phase diagram, including the relative positions of the solidus and liquidus boundaries and the composition dependence of the miscibility gap. Quantitatively, the liquidus deviations range from 2.82\% to 5.04\%, with the largest discrepancy at 90~mol\% NaCl. The solidus deviations range from 1.71\% to 9.36\%, again with the largest discrepancy at 90~mol\% NaCl. The mean deviations are 2.82\% for the liquidus and 5.02\% for the solidus. The larger errors near the NaCl-rich end are consistent with the upward shift observed for the pure NaCl melting point.

The predicted miscibility limits also follow the overall experimental composition dependence, although the quantitative agreement is less uniform. Deviations range from 0.20\% to 16.49\%, with the largest discrepancy near 30~mol\% NaCl and a mean deviation of 7.11\%. These results indicate that the model captures the broad temperature and composition trends associated with solid-phase mixing and demixing, but that the absolute phase boundaries retain non-negligible systematic errors.

Taken together, the phase-diagram calculations show that a single transferable MTP can reproduce the principal trends in NaCl--KCl phase behavior, including melting boundaries and the solid-phase miscibility gap. However, the comparison with experiment also highlights the limits of a purely first-principles-trained model for quantitative phase-boundary prediction. For engineering use, the most robust role of such a model may therefore be to interpolate trends across composition and temperature, while experimental data provide anchors for correcting systematic offsets in absolute transition temperatures and thermodynamic properties.

\subsection{Effect of the D3 dispersion correction}

To account for long-range dispersion effects in molecular dynamics simulations, we considered an MTP--D3 variant in which Grimme's D3 correction~\cite{grimme2010consistent} was added to the DFT-trained MTP. The underlying DFT reference data were generated without dispersion corrections. A single numerically converged D3 parameter set was used throughout, consisting of a 20~\AA\ cutoff with BJ damping. Results obtained with this dispersion-corrected variant are denoted as MTP--D3.

The influence of D3 was evaluated across structural, thermophysical, transport, and phase-equilibrium properties. As shown in Figs.~\ref{fig:latt-param}--\ref{fig:dif}, its effect is strongly system and property dependent. For crystalline lattice parameters obtained from static or near-static relaxation, the baseline MTP gives better agreement with experiment than MTP--D3 for NaCl, KCl, and several NaCl--KCl mixture compositions. For liquid densities, the effect of D3 is mixed: it substantially improves agreement for KCl, modestly improves NaK, has only a small effect for equimolar NaCl--KCl, and worsens agreement for NaCl. The radial distribution functions show little sensitivity to the D3 correction, which is consistent with the fixed-density protocol used for those calculations and indicates that the local liquid structure is reproduced similarly by both variants under the same volume constraint. For transport and phase-equilibrium properties, only MTP--D3 results are reported. Therefore, the direct effect of D3 on these observables cannot be isolated from the present data.

Overall, these comparisons show that D3 does not provide a uniformly transferable improvement across compositions or observables. The baseline MTP performs better for some properties and systems, whereas MTP--D3 improves agreement in others, particularly for KCl-rich or mixed liquid-density predictions. This variability indicates that the effect of empirical dispersion corrections depends on composition, thermodynamic state, and the property being evaluated. In the present system, the inclusion of D3 dispersion should be regarded as a physically motivated modeling variant, rather than as a correction that uniformly improves agreement with experiment.

More broadly, the results suggest that an MTP trained exclusively on DFT-PBE reference data can capture useful relative trends in molten alkali chloride systems, including temperature and composition dependencies in several thermophysical and phase-equilibrium properties. However, systematic offsets remain in some absolute quantities. These offsets are evident, for example, in the predicted melting temperatures, where deviations can approach $\sim$100~K depending on composition. Such discrepancies are important for engineering applications, even when the qualitative phase behavior and property trends are reproduced. MTP predictions based solely on DFT-PBE data should therefore be interpreted with care when absolute thermodynamic values are required.

These findings support a multi-scaling modeling strategy in which DFT-informed MLIPs can efficiently capture temperature- and composition-dependent behavior while maintaining physical consistency, thereby providing valuable predictive capability in areas of composition space that remain experimentally unexplored or sparsely characterized. However, achieving quantitative agreement with experiment may require calibration against selected experimental reference data. Such calibration can be introduced either during potential development through the inclusion of experimental targets or in downstream engineering workflows through property-level correction and validation. Integration of MTP-generated data into CALPHAD assessments, computational fluid dynamics (CFD) models, and materials-property databases should therefore account explicitly for possible systematic offsets to maintain quantitative accuracy, thermodynamic consistency, and predictive reliability.

\section*{Conclusion}

In this work, we developed a transferable MLIP based on the MTP framework for molten NaCl, KCl, NaCl--KCl mixtures, and the NaK system. The model was trained on DFT-PBE reference data generated using a small-cell active-learning strategy spanning a broad range of temperatures, compositions, and local atomic environments. We also considered an MTP--D3 variant, in which Grimme's D3 dispersion correction was added during molecular dynamics simulations using a consistent parameterization.

The results demonstrate that the treatment of dispersion interactions can significantly influence the predicted thermophysical properties; however, their inclusion does not consistently improve predictive accuracy across all properties and thermodynamic conditions. For some systems and properties, particularly KCl-rich and mixed liquid compositions, the D3 correction improves agreement with experiment. For others, including NaCl lattice parameters and molten NaCl density, the uncorrected MTP yields more accurate predictions. The effect of D3 is therefore system-, composition-, and property-dependent, and should be regarded as a physically motivated modeling variant rather than a correction that uniformly improve the results.

Across the systems studied, the MTP framework captures many of the key temperature- and composition-dependent trends in density, heat capacity, radial distribution functions, diffusion coefficients, thermal conductivity, and phase behavior. However, systematic deviations remain in several absolute properties. In particular, predicted melting temperatures can exhibit shifts approaching 100~K relative to experiment, depending on composition. These deviations are not simply uniform offsets, but composition-dependent biases that likely reflect limitations inherited from the underlying DFT reference data, the treatment of dispersion, finite-size effects, and the sensitivity of phase-boundary calculations.

Our results show that DFT-informed MTPs are powerful tools for exploring relative trends in molten-salt thermophysical and transport behavior, especially in composition and temperature regimes where experimental data are sparse. However, they should not be treated as fully quantitative predictors of absolute thermodynamic reference points without validation or calibration. For engineering applications, selected experimental anchors may be needed either during potential development, through augmented training objectives, or downstream, through integration with thermodynamic frameworks such as CALPHAD-type assessments.

Future work should therefore focus on systematically assessing the influence of exchange--correlation functional, dispersion treatment, training-set construction, and hybrid first-principles/experimental calibration strategies. Extending this approach to more complex multicomponent molten-salt systems will be essential for building reliable predictive workflows. Overall, this study demonstrates that DFT-informed MTPs provide an efficient framework for capturing composition- and temperature-dependent behavior in molten salts, while also clarifying the need for experimental anchoring when quantitative absolute accuracy is required.

\section*{Acknowledgements}
\noindent We thank the Digital Research Alliance of Canada (DRAC) for generous allocation of compute resources. Financial support was provided by the Natural Sciences and Engineering Research Council of Canada (NSERC), Mitacs, and the University Network for Excellence in Nuclear Engineering (UNENE). This work was partly funded by Atomic Energy of Canada Limited, Canada, under the auspices of the Federal Nuclear Science and Technology Program.

\section*{Declaration of generative AI and AI-assisted technologies in the writing process}

Generative AI (ChatGPT 5) was used to assist with English-language editing, formatting, grammar checking, and refinement of the original manuscript drafts. All algorithms and data-processing methods were conceived and developed by the authors. ChatGPT was used only to assist with the implementation, optimization, and refinement of the corresponding Python code to improve coding efficiency and streamline the workflow. The authors confirm that all data, results, figures, analyses, and scientific interpretations were conceived, produced, and verified by the authors and were not generated by AI.

\section*{Data availability}
All Python scripts developed for this work, together with the datasets and machine-learning interatomic potentials used in this study, are publicly available at: \url{https://github.com/kzongogith/NaCl-KCl-Phase-Diagram-MTP-D3}
\section*{Author contributions}
\noindent L.K.B. initiated and coordinated the research project and provided the necessary computing resources. K.Z. performed all DFT calculations, trained the MTP, conducted molecular dynamics simulations, and carried out the analyses presented in this work. All authors—K.Z., H.S., E.T., C.M., and L.K.B.—contributed to writing the manuscript.
\section*{Competing interests}
\noindent The authors declare no conflict of interest.

\bibliographystyle{unsrt}
\bibliography{ref}

\end{document}